\documentclass[11pt]{article}
\usepackage{moriond,epsfig}

\def\be{\begin{equation}}
\def\ee{\end{equation}}
\def\bea{\begin{eqnarray}}
\def\eea{\end{eqnarray}}

\begin{document}
%%%% moriond setting: \vspace*{4cm}
\vspace*{3cm}
\title{FLAVOR VIOLATION IN SUSY}
DO-TH 09/06
\author{ G. HILLER }

\address{Institut f{\"u}r Physik, Technische Universit{\"a}t Dortmund,
  D-44221
  Dortmund, Germany}

\maketitle\abstracts{I review recent opportunities from flavor 
for $b$-physics and collider searches. }

\section{Introduction: Flavor physics}

The starting point for flavor physics is that known matter comes in generations 
 $\Psi_i$, $i=1,2,3$, where each generation carries identical
 charges under the Standard Model (SM) gauge group.
 Within the SM, the generations are solely distinguished by the Yukawa interactions, 
 which couple the fermions to the Higgs boson. The non-diagonal structure of the
 Yukawa matrices allows for quark mixing and flavor change under the weak interaction. 
 The strength of the transition of (quark) flavor $i  \to j$ is encoded
 in the CKM mixing matrix element $V_{ji}$. In matrix form:
 \begin{equation}
V= \left( \begin{array}{lll}
V_{ud}  &  V_{us} & V_{ub}\\
V_{cd}  &  V_{cs} & V_{cb}\\
V_{td}  &  V_{ts} & V_{tb}
\end{array} \right) \sim
\left( \begin{array}{lll}
1 &  \lambda &  \lambda^3\\
-\lambda  &  1 & \lambda^2 \\
-\lambda^3 &  -\lambda^2 & 1
\end{array} \right) ,
 \end{equation}
 where the phenomenological parametrization in terms of the Wolfenstein parameter 
 $\lambda \simeq 0.22$ reflects the strong hierarchy present in the quark mixing.
The third generation ($t,b$) is decoupled from the first two at order
$\lambda^2$.
Today, masses and mixings including CP-violation are known input,
with still improving accuracy through experimental and theoretical efforts. The origin of the hierarchical quark mass
pattern and mixing, however, remains unexplained within the SM.
 
Flavor changing neutral current (FCNC) processes are suppressed in the SM by 
several effects: they arise only at the loop level, they are suppressed by flavor mixing 
and, except for the top quark contribution, by small mass splittings  "GIM-mechanism".
Since existing FCNC data are consistent with the SM, 
a mechanism to control FCNCs is required for any SM extension.

There are several possibilities and combinations thereof:
{\it i} There is no flavor suppression from contributions beyond the SM to FCNCs. They are small
because  the scale of New Physics (NP) is high  and not connected to the scale of electroweak symmetry breaking, hence, not in reach of the Large Hadron Collider (LHC).
Since much of our reasoning about extending the SM is driven by electroweak symmetry breaking, we will not consider this any further and assume a TeVish NP scale.
{\it ii}  There is no (significant) flavor suppression but we allow for contributions to cancel to
reproduce the data. Since not every sensible observable is measured yet (with sufficient precision) there is the intriguing possibility that the SM breaks down in the next round of experiments
\cite{Buchalla:2008jp,superb}. Especially promising
here are rare decays and $B_s -\bar B_s$-mixing.
{\it iii} There is a flavor suppression in the NP contributions similar to the one of the SM, which is
$ \lambda^n$. Models with the same amount of flavor violation as the SM with the only
source of flavor being the Yukawa matrices, are termed
Minimal Flavor Violation (MFV)-type \cite{mfv,D'Ambrosio:2002ex}.  We discuss solutions to the flavor problem and their signatures for the case of supersymmetry (SUSY) below.

\section{Flavor in SUSY}

The superpotential of the minimal supersymmetric standard model (MSSM) with
unbroken $R$-parity, neglecting leptons, reads as
$W_{MSSM}=Q_i (Y_U)_{ij} H_u \bar U_j+ Q_i (Y_D)_{ij} H_d \bar D_j -\mu H_d H_u$ .
It contains no further sources of flavor violation than the SM, that is, the up and down quark Yukawas, $Y_U$ and $Y_D$.  Its supersymmetric contribution to the MSSM Lagrangian is therefore always MFV.
Whether the full model is MFV,  depends then on the breaking of supersymmetry.
%(we assumed no flavor  distinction in the kinetic terms).

Within MFV, the
SUSY-breaking greatly simplifies  \cite{D'Ambrosio:2002ex}, here for $SU(2)$-doublet masses,
as \cite{Allanach:2009ne}
\begin{eqnarray}  
({m}_{\tilde Q}^{2})^T &=& z^q_{1}\,\mathbf{1}+z^q_{2} {Y}_{U}
{Y}_{U}^{\dagger}+z^q_{3} {Y}_{D}{Y}_{D}^{\dagger}+z^q_{4}
({Y}_{U}{Y}_{U}^{\dagger})^{2}+z^q_{5}({Y}_{D}
{Y}_{D}^{\dagger})^{2}  +\ldots 
\, ,
\label{eq:mq2g} 
\end{eqnarray}
up to terms involving higher powers of the Yukawas and analogous expressions for singlets and the trilinear terms. For $z_{i \neq1} \equiv 0$ the soft terms are called flavor blind.
Well-known MFV models are gauge and anomaly mediation, and by construction, mSUGRA and the CMSSM. The constrained flavor structure of the MFV-type MSSM yields generic predictions:
 Highly degenerate squarks of 1st and 2nd generation and a 3rd generation decoupled
at order $\lambda^2$.
\begin{figure}
\begin{center}
%\rule{5cm}{0.2mm}\hfill\rule{5cm}{0.2mm}
%\vskip 2.5cm
%\rule{5cm}{0.2mm}\hfill\rule{5cm}{0.2mm}
\vspace{-0.4cm}
\psfig{figure=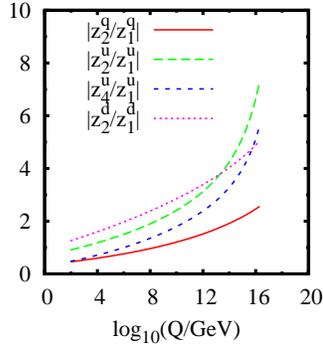,height=2.0in}
\vspace{-0.4cm}
\end{center}
\caption{RG evolution of MFV coefficients within AMSB for $\tan \beta=10$.
Figure taken from Ref.~5.
\label{fig:RGE}}
\end{figure}

Renormalization group (RG) running modifies all MFV coefficients $z_i$. In particular, a 
model flavor blind at the scale of SUSY breaking such as gauge mediation develops squark masses with non-trivial flavor structure at the weak scale \cite{Paradisi:2008qh}.  While also being MFV, the behavior of
anomaly mediated SUSY breaking (AMSB) is of marked difference: 
The amount of flavor violation decreases from the high to the weak scale,
see Fig.~\ref{fig:RGE}, where we show ratios $|z_{i>1}/z_1|$ as a function of the scale. Moreover, for low $\tan \beta$,  AMSB becomes exactly flavor blind
in the quasi-infrared fixed point limit of the top Yukawa \cite{Allanach:2009ne}.

\section{Probing squark flavor}

How to experimentally identify MFV and non-MFV variants ?

\subsection{With $b$-physics}

\begin{figure}
\begin{center}
%\rule{5cm}{0.2mm}\hfill\rule{5cm}{0.2mm}
%\vskip 2.5cm
%\rule{5cm}{0.2mm}\hfill\rule{5cm}{0.2mm}
\vspace{-0.5cm}
\psfig{figure=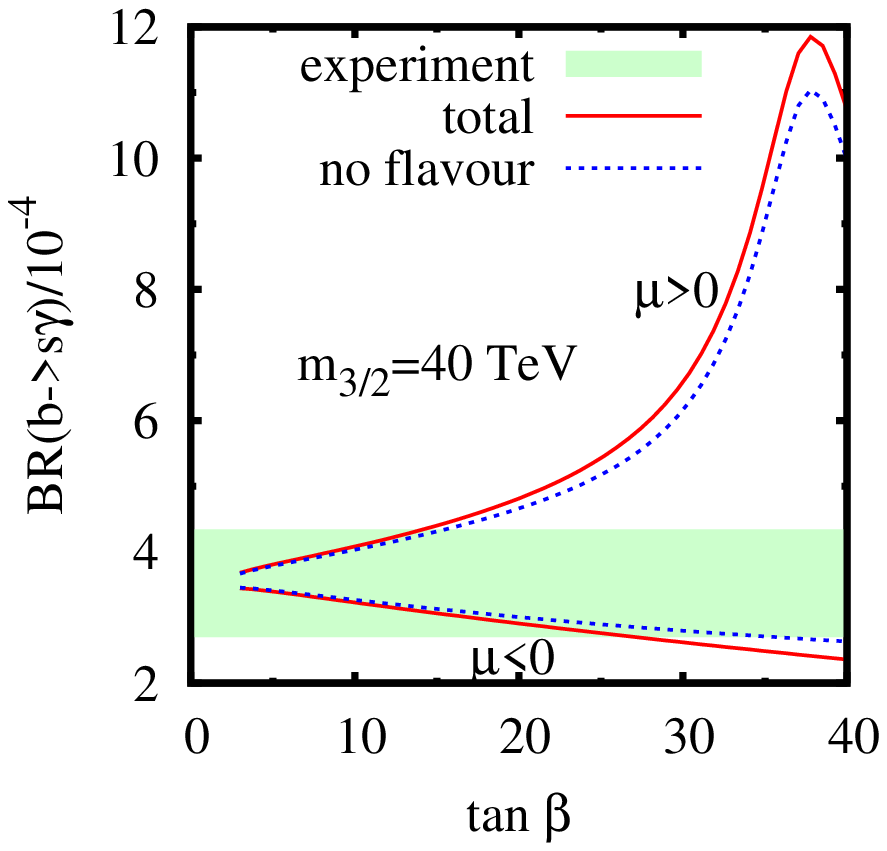,height=2.0in}
\hspace{-1.0cm}
\psfig{figure=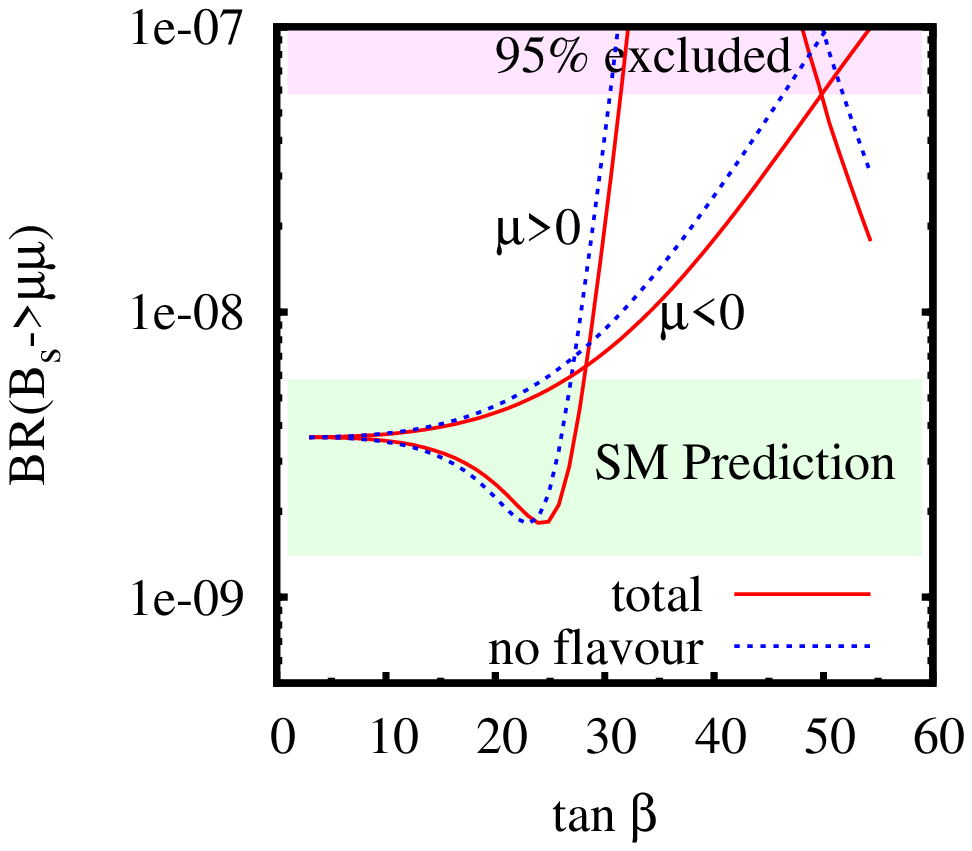,height=2.0in}
\vspace{-0.4cm}
\end{center}
\caption{Rare decays within mAMSB with or without flavor for either sign of $\mu$. 
Left: $b\to s \gamma$, right: $B_s \to \mu \mu$.
\label{fig:bs}}
\end{figure}

It is well-known that within MFV, ${\cal{O}}(1)$  effects are possible in rare processes if $\tan \beta$ is largish. With its UV insensitivity, AMSB is particularly predictive.
Analytical expressions for the full flavor structure, that is, the $z_i$ and the "mass insertions"
$(\delta)_{ij}$ exist \cite{Allanach:2009ne}. AMSB predictions for the branching ratios of
$b \to s \gamma$ and $B_s \to \mu^+ \mu^-$ decays are shown in Fig.~\ref{fig:bs}.
The leptonic mode will place additional constraints on $\tan \beta$ (depending on the gravitino mass $m_{3/2}$) for $\mu<0$ once the upper bound on its branching ratio improves.
Since in ASMB the squarks are generically  heavy, the impact
of a flavor universal shift to avoid tachyonic sleptons on squark flavor is small 
\cite{Allanach:2009ne}. 

Deviations from the SM can be more dramatic in non-MFV models.
Key indicators are right-handed FCNCs, the presence of non-CKM CP-violation or
the breakdown of relations between observables linked by CKM. Recently, a lot of theory interest
has focussed on angular distributions in $B \to (K^* \to K \pi) \mu^+ \mu^-$ 
\cite{Bobeth:2008ij} \cite{Egede:2008uy},  also 
$B \to K   l^+ l^-$  \cite{Bobeth:2007dw} decays.
Opportunities arise from CP-asymmetries in 
semileptonic $B \to K^*$ and $B_s \to \Phi$ transitions, which are
doubly-Cabibbo suppressed within MFV (and the SM) and sensitive to a multitude of NP couplings 
\cite{Bobeth:2008ij}: 
Four CP-asymmetries are CP-odd, hence, can be extracted without flavor tagging.
This is advantageous for  $\bar B_s, B_s \to (\Phi \to K K) \mu^+ \mu^-$ decays.
Three CP-asymmetries are odd under  $T_N$, the naive T-transformation. They can be
order one with NP even if the strong phases are small such as predicted in the framework of QCD factorization for low dilepton invariant mass.

\subsection{At ATLAS/CMS: Long live the stop}

\begin{figure}
\begin{center}
%\rule{5cm}{0.2mm}\hfill\rule{5cm}{0.2mm}
%\vskip 2.5cm
%\rule{5cm}{0.2mm}\hfill\rule{5cm}{0.2mm}
\vspace{-0.2cm}
\psfig{figure=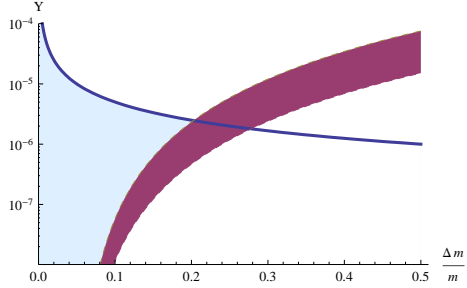,height=1.5in}
\vspace{-0.2cm}
\end{center}
\caption{The $\tilde t c \tilde \chi^0$ coupling $Y$ versus  $\Delta m/m_{\tilde t}$. The stop has macroscopic decay length below the solid curve and decays dominantly through FCNCs to the left of the dark band. Figure adopted from Ref.~10.
\label{fig:Yvsdelm}}
\end{figure}

It is an interesting question whether MFV and its CKM-like squark mixing can be measured at colliders. In MFV, mixing between the stop and first two generations is suppressed
$ (\tilde m^2_{Q})_{23}/\tilde m^2 \sim y_b^2V_{cb}V_{tb}^* \sim 10^{-5} 
\tan \beta^2$, see Eq.~(\ref{eq:mq2g}). Such a tiny coupling
can nevertheless be probed if  $\tilde t \to c \chi^0$ is the dominant decay of the lightest stop 
$\tilde t$ and its rate is sufficiently suppressed  \cite{Hiller:2008wp}.
Then, 
\begin{equation}
\tau_{\tilde t}\sim\ {\rm ps}\ \left(\frac{m_{\tilde t}}{100\ \mbox{GeV}}\right
)
\left(\frac{0.03}{\Delta m/m_{\tilde t}}\right)^2\left(\frac{10^{-5}}{Y}\right)^2  , 
~~~\Delta m=m_{\tilde t}-m_{\chi^0}  , ~~~Y_{\rm MFV} \sim y_b^2 V_{cb} V_{tb}^*  ,
\end{equation}
the lifetime of the stop is long and yields a macroscopic decay length. The flavor diagonal
decay of the stop into top can be forbidden kinematically by having the stop-neutralino splitting 
$\Delta m$ below the top mass.
For $\Delta m > m_b$, also the 4-body  decays $\tilde t \to b \chi^0 l \nu$ open up.
Approximately,
\begin{equation}
\frac{\Gamma(\tilde t\to b\chi^0 l\nu)}{\Gamma(\tilde t\to c\chi^0)}
\approx\frac{g^6|V_{tb}|^2}{2 (4 \pi)^4 }\frac{(\Delta m-m_b)^8}
{[Y (\Delta m)]^2 m_W^4 m_{\chi^+}^2}  .
\end{equation}
Relevant regions of stop parameters are shown in Fig.\ref{fig:Yvsdelm} for $m_{\tilde t}=100$ GeV. The stop has decay length  $\beta \gamma \tau_{\tilde t} >0.1$ mm and dominant FCNC decay in the light shaded (blue) region. Experimental constraints exist for larger values of $
\Delta m$ \cite{Abazov:2008rc}.
Observation of a long-lived stop would strongly support MFV since  other scenarios have
generically much larger $\tilde t c \tilde \chi^0$ couplings $Y$  \cite{Hiller:2008wp}.
A light stop is ingredient of electroweak baryogenesis and supports coannihilation of the relic density \cite{Boehm:1999bj}; it  can be realized in hypercharged anomaly mediation \cite{Dermisek:2007qi}, or with large  trilinear $A$-terms. 

\section{Summary}

What are the flavor quantum numbers of the new particles and SM partners related to
electroweak symmetry breaking ?
Already strong constraints exist from $K,D$- and $B$-observables,  supporting
models with flavor suppression not too far from the one in the SM, perhaps even exactly as in the SM, called MFV.  
The LHCb experiment and possibly super flavor factories \cite{superb}  will greatly contribute to clarifying this question, but also the ATLAS/CMS experiments, as the possibility  to learn about 
squark flavor mixing  from a stop decay length measurement
shows.

\section*{Acknowledgments}
G.H. is  happy to thank her collaborators, and the organizers for the opportunity to come and
speak at Moriond EW 09. The work reported here is supported in
part by the Bundesministerium f\"ur Bildung und Forschung, Berlin-Bonn,
and the German-Israeli-Foundation (G.I.F.).

\end{document}